# Cooling power analysis of a small scale 4 K pulse tube cryocooler driven by an oil-free low input power Helium compressor

J-A Schmidt[1,2], B Schmidt[1,2], J Falter[2], J Höhne[3], C Dal Savio[4], S Schaile[4], A Schirmeisen[1,2]

[1]Justus-Liebig-University Giessen, Germany
[2]TransMIT GmbH, Giessen, Germany
[3]Pressure Wave Systems GmbH, Taufkirchen, Germany
[4]attocube systems AG, Haar, Germany

**Abstract.** Here we report the performance of a small scale 4 K pulse tube cryocooler operating with a low input power reaching a minimum temperature of 2.2 K, as well as a cooling capacity of over 240 mW at 4.2 K. The compressor is air cooled and can be supplied by single phase power sockets. With an input power of about 1.3 kW the coefficient of performance reaches values of up to 185 mW/kW, which is among the highest currently reported values for small to medium power pulse tubes. The combination of an oil-free Helium compressor and low maintenance pulse tube cryocooler provides a unique miniaturized, energy efficient and mobile cooling tool for applications at 4 K and below.

**1. Introduction**
In the last decade the field of quantum technologies experienced a strong increase in interest [1, 2], which at the same time spurred developments in cryocooling technologies. Apart from large quantum computer systems, there is a fast-growing demand for small low power, highly mobile and minimum maintenance cooling systems at 4 K for optical quantum components, e.g., single photon detectors, sensor arrays or single photon emitters and small superconductive electronic circuits. Due to increased energy costs and sustainability issues, energy efficiency of the cooling systems is becoming more important. A typical transmitter for quantum signals generates only low electrical heat but needs to be cooled to temperatures of 4 K and below for low noise operation. Therefore, a closed-cycle small scale cooling system with compact size that allows for easy transport and installation is advantageous [3]. Recently the smallest currently available 4 K pulse tube cryocooler SUSY was presented, which achieved 140mW cooling power at 4.2 K running with a conventional 1kW oil capsule based helium compressor [4]. At the same time a new technology for oil-free Helium compressors was developed, based on a mechanism with hydraulically driven metal bellows [5]. The air-cooled variable power IGLU compressor provides minimum maintenance and maximum mobility with a miniature footprint and can be supplied by single phase power sockets. Additionally, the compressor offers to be adjusted in terms of input power to reduce energy consumption e.g. for standby application. The combination of a small scale cryocooler with low maintenance, and low input power Helium compressor IGLU [3,4] is an ideal candidate for future small scale quantum applications [4, 5]. Here we present the first lab results of the combined systems in terms of cooling power and coefficient of performance with two driving speeds of the compressor.

## 2. Experimental setup

The single phase IGLU compressor (Figure 1a) is air-cooled with an input power range of 0 to 1.3 kW. It is 19" rack compatible and can be operated in any orientation. The system can be remotely adjusted in power and cold head speed for standby operation. The attached pulse tube cryocooler SUSY (Figure 1b) has been developed for the input power range of around 1 to 2 kW and is remotely connected to and electrically insulated from the rotary valve. The connecting metal hoses between cryocooler and compressor are 10 m long with an inner diameter of DN12.

The cryocooler is put in a vacuum vessel with sufficient vacuum conditions of $<1\cdot 10^{-4}$ mbar. The second stage is covered with a Copper radiation shield, which is connected to the first stage of the cryocooler. Both stages and the shield are wrapped in several layers of reflective Mylar foil to reduce thermal heat load. The first and second stage are equipped with electrical heaters to simulate heat loads generated by application, i.e., optical quantum components.

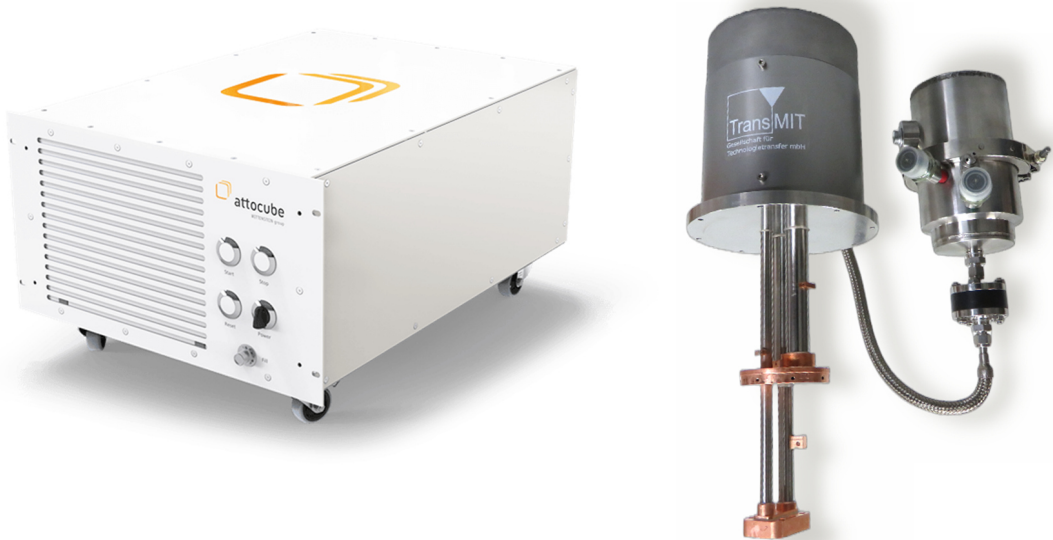

**Figure 1**. (a) Picture of the IGLU Helium compressor with variable input power. (b) Pulse tube cryocooler cold head PTD SUSY with remote rotary valve.

## 3. Cooling performance

All performance data was taken following the same methodology, as described in the previous section. The cooldown of the combined system is displayed in Figure 2. The first stage temperature reaches a minimal temperature of <45 K after 150 min, while the second stage achieves a temperature below 4.2 K after 85 min and a minimal temperature of <2.2 K. During the cooldown the IGLU compressor was operated at maximum speed with an electrical input power of 1.3 kW.

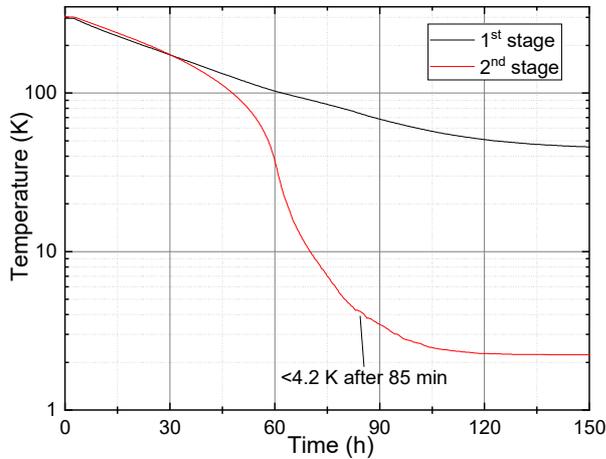

**Figure 2**. Measured cooldown with IGLU compressor and SUSY pulse tube. The first stage (black line) achieves a minimal temperature of <42 K and the second stage temperature is below 4.2 K after 85 min (red line). Input power of the IGLU compressor was 1.3kW.

After the cooldown period the cooling capacity maps were measured for different heating powers at the first and second stage, as displayed in Figure 3a. The compressor can be operated at different speeds, resulting in different Helium flux levels. The average operating pressure was 19.2 bar, and the pressure difference between high and low pressure was 0.85 MPa. In the capacity maps experiments the compressor was set to the minimum and maximum speed, resulting in an electrical input power of 900 W (black squares) and 1.3 kW (red circles), respectively. Different electric heat loads were applied simultaneously at the first and second stage of the cryocooler. The heat load applied at the electrical resistive heaters was in the range of 0 to 2 W for the first stage and 0 to 0.3 W for the second stage.

The cooling capacity maps are markedly different for the low and high input power levels of the compressor, as expected. For zero heat load at the second stage, the first stage reaches 41 K for the high power level, which increases to 49 K at the low input power setting. More importantly, with zero heat load at the first stage, the low temperature second stage has a cooling capacity of 240 mW at 4.2 K for the high power setting. At the low power compressor setting the system still achieves 140 mW at 4.2 K, which is still well above the 100 mW level. For both input power levels, the SUSY cooler reaches the physical limit of 2.2 K when no heat loads are applied.

For optical quantum sensor applications, where the superconducting transition temperature is critical, the cooling capacity at 3 K is of particular importance. In this test configuration we reached 80 mW and 55 mW for the high and low power compressor operation, respectively. These 2$^{nd}$ stage cooling capacities are influenced by the 1$^{st}$ stage temperatures, as expected. However, the typical 'dip' of 2$^{nd}$ stage cooling capacity for increased 1$^{st}$ stage temperature, often observed for conventional pulse tubes, is not found for this miniature SUSY pulse tube. In fact, the 2$^{nd}$ stage heating capacity rather decreases in a linear fashion with the 1$^{st}$ stage temperature.

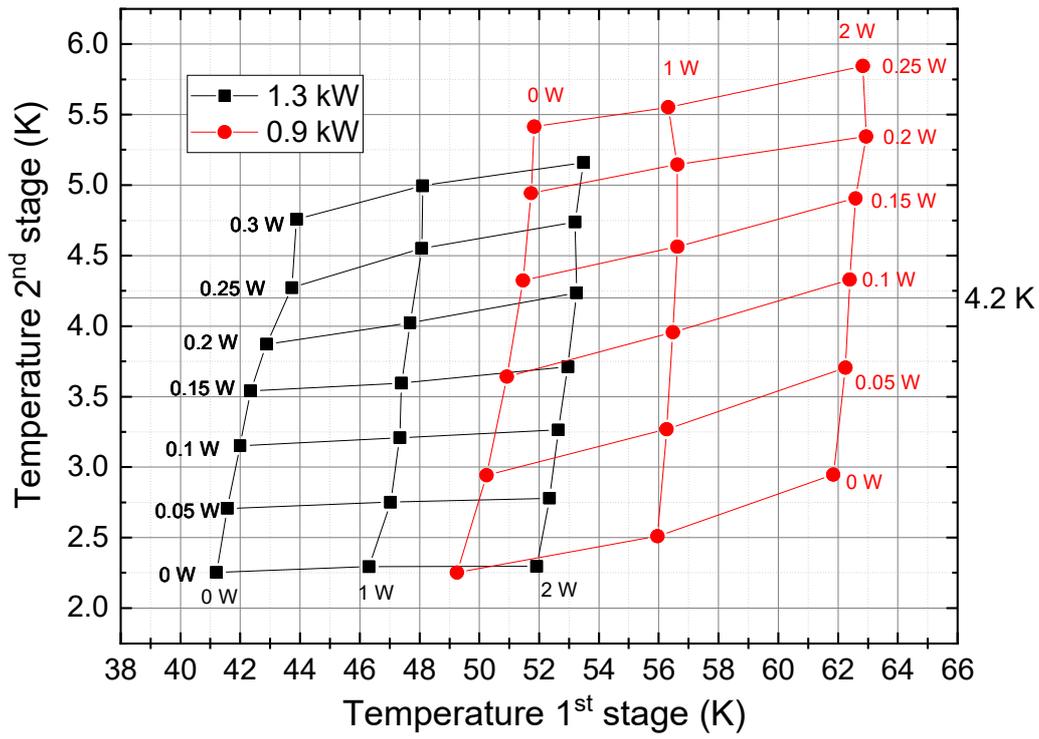

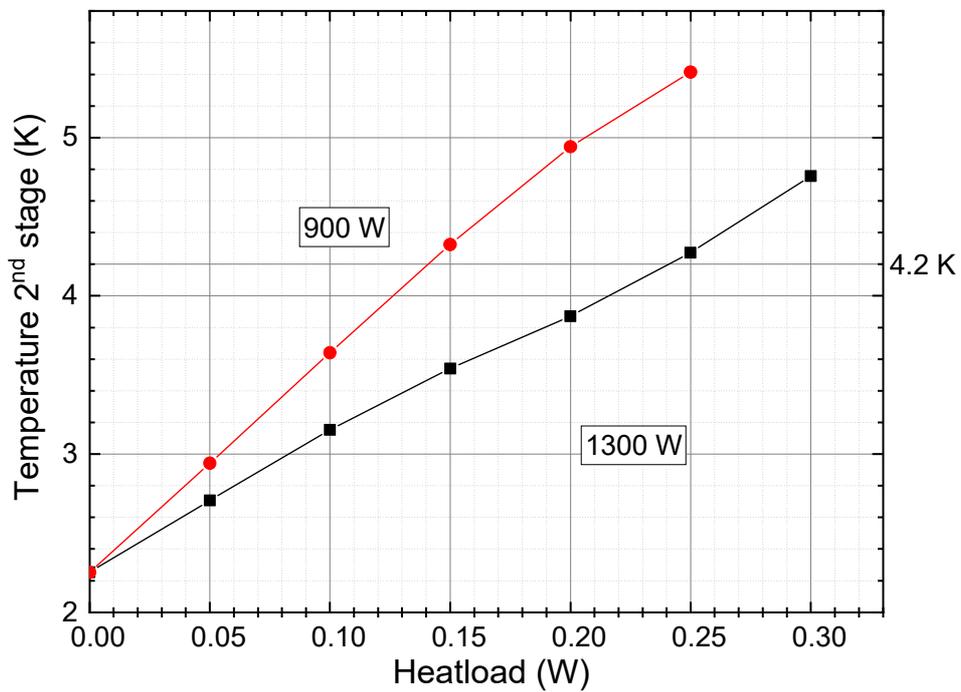

**Figure 3**. (a) Measured capacity map of the combined system of pulse tube SUSY and Helium compressor IGLU with an electrical input power of 900 W (red circles) and 1.3 kW (black squares). (b) Load curve of second stage without heat load applied at first stage for 900 W (red circles) and 1.3 kW (black circles) electrical input power.

In Figure 3b a close up of the 2nd stage temperature as a function of the 2nd stage heat load is displayed, with no load at the first stage. This visualizes the almost linear relation between 2nd stage temperature and heat load over a wide range. We find slopes of 1 K and 1.3 K per 100mW heat load for the high and low power compressor setting, respectively. Finally, we quantify the cryocoolers coefficient of performance, defined as the ratio between the cooling power at 4.2 K and the electrical input power of the compressor. This COP reaches values of 185mW/kW for the high input power setting. In the class of low to medium sized compressors, with input powers from 1 kW to 10 kW, this is the highest so far reported COP of a pulse tube compressor combination, where the average COPs typically range from 40 to 160 mW/kW. For the low input power setting of IGLU the COP decreases to 155 mW/kW, which is still among the highest efficiency values. Please note that this is achieved in the remote configuration, i.e., the rotary valve is separated from the pulse tube head by a pressure hose of 25 cm to decouple vibrations. We anticipate even better COP performance for direct coupling of the valve to the head, when vibration decoupling is not application relevant.

**4. Conclusion**

We have presented the combination of a low input power Helium compressor and small scale pulse tube cryocooler. The cryocooler achieves minimum temperatures down to <2.2 K, while the cooling power reaches 240 mW at 4.2 K at 1.3 kW compressor input power, as well as up to 80 mW cooling power at 3 K. The adjustable compressor offers to reduce its electrical input power, which results in an available cooling power of still 140 mW at 4.2 K with 900 W. This combination is also energy efficient with a COP of 185 mW cooling power per 1 kW input power, placing this combination at the top of currently available cryocoolers in the low to medium input power range.

## 5. Acknowledgments
We thank Yusuf Kuecuekkaplan for his excellent support, as well as ongoing fruitful discussions with Prof. Günter Thummes.
## 6. References
[1]     S Golestan, MR Habibi, SY Mousazadeh Mousavi, JM Guerrero and JC Vasquez 2023 Energy Reports 9
[2]     S Sikiru, TL Oladosu, SY Kolawole, LA Mubarak, H Soleimani, LO Afolabi and AOO Toyin 2023 Journal of Energy Storage 60
[3]     J O'Brien, A Furusawa, and J Vučković Photonic quantum technologies 2009 Nature Photon 3
[4]     JA Schmidt, B Schmidt, D Dietzel, J Falter, G Thummes and A Schirmeisen Cryogenics 122
[5]     J Höhne 2014 AIP Conference Proceedings 1573
[6]     Ray Radebaugh 2009 J. Phys.: Condens. Matter 21

## 5. Acknowledgments
We thank Yusuf Kuecuekkaplan for his excellent support, as well as ongoing fruitful discussions with Prof. Günter Thummes.

## 6. References
[1]     S Golestan, MR Habibi, SY Mousazadeh Mousavi, JM Guerrero and JC Vasquez 2023 Energy Reports 9
[2]     S Sikiru, TL Oladosu, SY Kolawole, LA Mubarak, H Soleimani, LO Afolabi and AOO Toyin 2023 Journal of Energy Storage 60
[3]     J O'Brien, A Furusawa, and J Vučković Photonic quantum technologies 2009 Nature Photon 3
[4]     JA Schmidt, B Schmidt, D Dietzel, J Falter, G Thummes and A Schirmeisen Cryogenics 122
[5]     J Höhne 2014 AIP Conference Proceedings 1573
[6]     Ray Radebaugh 2009 J. Phys.: Condens. Matter 21